\documentclass[aps,prx,reprint,noeprint,floatfix]{revtex4-2}
\usepackage[utf8]{inputenc}
\usepackage{xcolor}
\usepackage{amsmath}
\usepackage{amssymb}
\usepackage{physics}
\usepackage{graphicx}
\usepackage{parskip}
\usepackage{hyperref}
\usepackage{varioref}       
\usepackage{cleveref}     

\crefname{table}{table}{tables}
\Crefname{table}{Table}{Tables}
\crefname{figure}{fig.}{figs.}
\Crefname{figure}{Fig.}{Figs.}
\crefname{section}{sect.}{sects.}
\Crefname{Section}{Sect.}{Sects.}

\definecolor{nicecolor}{rgb}{0.03, 0.27, 0.49}
\hypersetup{colorlinks=true,allcolors = nicecolor,linktocpage=true}

\usepackage{mathptmx}

\newcommand{\commie}[1]{}

\newenvironment{eqaed}
    {\begin{equation}
    \begin{aligned}
    }
    { 
    \end{aligned}
    \end{equation}
    \ignorespacesafterend
    }

\begin{document}

\begin{titlepage}

\begin{flushright}
LMU-ASC 13/24 \\
MPP-2024-175
\end{flushright}

\title{\vspace{-32pt}\bf Dark dimension with (little) strings attached}
\author{Ivano Basile}
\email{ibasile@mpp.mpg.de, ivano.basile@lmu.de}
\author{Dieter L\"{u}st}
\email{luest@mpp.mpg.de, dieter.luest@lmu.de}
\affiliation{Arnold-Sommerfeld Center for Theoretical Physics\\ Ludwig Maximilians Universit\"at M\"unchen\\ Theresienstraße 37, 80333 M\"unchen, Germany}
\affiliation{Max-Planck-Institut f\"ur Physik (Werner-Heisenberg-Institut)\\ Boltzmannstraße 8, 85748 Garching, Germany}

\begin{abstract}

    \noindent We motivate a relation between dark energy and the scale of new physics in weakly coupled string theory. This mixing between infrared and ultraviolet physics leads to a unique corner for real-world phenomenology: barring fine-tunings, we are naturally led to the ``dark dimension'' scenario, a single mesoscopic extra dimension of micron size with the standard model localized on D-branes. Our explicit top-down worldsheet derivation establishes it on a more solid grounding. Allowing some fine-tuning, such that the vacuum energy only arise at higher orders in string perturbation theory, the ``little string theory'' scenario with a very weakly coupled string is an alternative possibility. In this case, the string scale lies at the edge of detectability of particle accelerators.
    
    \end{abstract}

\maketitle

\end{titlepage}
\setcounter{page}{1}
\section{Introduction}\label{sec:introduction}

The paradigm of effective field theory (EFT) has been extremely successful in describing all aspects of physics and incorporating our ignorance of more microscopic details in terms of irrelevant parameters. In particle physics and quantum gravity, the spectacular successes of this approach can be ascribed to the fact that, at low energies, the (renormalizable sector of the) standard model (SM) and general relativity (GR) are accompanied by irrelevant corrections which are expected to be dramatically suppressed, leading to the powerful framework of GR-SM-EFT. Whatever mass scale $m_\text{gap}$ gaps new physics in the matter sector has eluded experimental searches thus far. Genuinely gravitational effects beyond the EFT arise at a potentially different scale $\Lambda_\text{UV}$, which also remains mysterious. At the EFT level, there is no way of predicting these quantities \footnote{The gravitational cutoff $\Lambda_\text{UV}$ is (bounded above by) the \emph{species scale} \cite{Veneziano:2001ah, Dvali_2010, Dvali:2007wp, Dvali:2010vm, Castellano:2023aum, Basile:2023blg, Basile:2024dqq, Bedroya:2024ubj, Aoufia:2024awo}.}.

Intriguingly, this is not the whole story. A markedly low-energy scale which still puzzles physicists to this day is embodied by dark energy \footnote{We do not assume that dark energy be a cosmological constant.}. In terms of the Planck scale $M_\text{Pl} \approx 10^{19} \text{ GeV}$, dark energy is observed to have the astonishingly small positive value $\Lambda_\text{dark} \approx 10^{-120} \, M_\text{Pl}^4$. Instead of predictability, the mystery of dark energy lies in its apparently unnatural smallness, which cannot be explained in an EFT framework.

Naively, the lowest and highest energy scales in physics have nothing in common: the Wilsonian paradigm of decoupling places an enormous chasm between them. Going beyond GR-SM-EFT, the most important lesson of quantum gravity is that this naive expectation is not correct. Apparently unrelated scales can be connected by non-trivial relations mixing infrared (IR) and ultraviolet (UV) physics. This observation dates back to early developments on black-hole thermodynamics, and is the conceptual foundation of the swampland program \cite{Vafa:2005ui, Palti:2019pca, Agmon:2022thq}. This notion is corroborated by string theory, where remarkable relations of this type commonly arise. On general bottom-up grounds \cite{Stout:2022phm} and from such top-down examples, it has been suggested that small dark energies are linked to a small gap to an infinite tower of new species \cite{Lust:2019zwm} by scaling relations of the form
\begin{eqaed}\label{eq:scaling_relation}
    \Lambda_\text{dark} \sim m_\text{gap}^\alpha
\end{eqaed}
up to a prefactor, for some positive $\alpha$ which remains $\mathcal{O}(1)$ in the limit under consideration. Unless otherwise stated for emphasis, here and in the following we set $M_\text{Pl} = 1$. The scale $\Lambda_\text{UV}$ of new gravitational physics is also generically linked to $m_\text{gap}$ by a scaling similar to \cref{eq:scaling_relation}, leading to small $\Lambda_\text{UV}$ as well. These ideas have been combined with observational data to suggest the existence of a mesoscopic extra dimension of roughly micron size, the ``dark dimension'' \cite{Montero:2022prj}.

In this letter we derive and discuss a relation of this type from string theory. More precisely, in closed weakly coupled string theory with four (extended) spacetime dimensions we find
\begin{eqaed}\label{eq:punchline}
    \Lambda_\text{dark} \sim m_\text{gap}^4 \, .
\end{eqaed}
Combining the stringent consistency requirements of string theory with observational data, \cref{eq:punchline} leads to two possible scenarios for phenomenology:

\begin{enumerate}

   \item With higher-dimensional supersymmetry and without further fine-tunings, the dark dimension scenario of \cite{Montero:2022prj}, with a single mesoscopic extra dimension of roughly micron size. In this case, $m_\text{gap} = m_\text{KK} = \Lambda_\text{dark}^{\frac{1}{4}} \approx 10 \text{ meV}$ is the Kaluza-Klein gap and the string coupling $g_s \approx 10^{-2}$ is related to SM gauge couplings.
   
   \item Assuming some fine-tuning with vanishing one-loop contribution, the ``little string theory'' scenario of \cite{Antoniadis:2001sw}, where strings are resolvable around accelerator energies $E_\text{acc} = \Lambda_\text{dark}^{\frac{1}{8}} \approx 10 \text{ TeV}$ \cite{Lust:2008qc, Anchordoqui:2008di}. Here $m_\text{gap} = M_s = E_\text{acc}$ is the string scale, and $g_s = \Lambda_\text{dark}^{\frac{1}{8}} \approx 10^{-15}$ is not tied to SM gauge couplings, which arise from NS5-branes.

\end{enumerate}

In the remainder of the letter we motivate this conclusion, providing a sketch of the derivation. A complete derivation with details and examples will be provided in a companion paper.

\section{Derivation from the worldsheet}\label{sec:worldsheet_derivation}

Aside from its appeal as a unified theory of gravity and gauge interactions, weakly coupled string theory in four (extended) spacetime dimensions is a suggestive phenomenological setting also from the bottom-up perspective of the swampland program. Indeed, the optimistic case $m_\text{gap} \, , \, \Lambda_\text{UV} \ll M_\text{Pl}$ implies that gravity is weakly coupled, since the quantum couplings of gravitons at the cutoff are captured by $\frac{\Lambda_\text{UV}^2}{M_\text{Pl}^2} \ll 1$. In addition to the obvious advantages for future experiments, this scenario may be mandated by a consistent parametric control over Eguchi-Hanson instantons \cite{Dvali:2024dlb}. One way for a theory to achieve this is to have super-Planckian extra dimensions, with $m_\text{gap} = m_\text{KK} \ll M_\text{Pl}$ and $\Lambda_\text{UV} \ll M_\text{Pl}$ the higher-dimensional Planck scale. This case remains within the bounds of EFT; the alternative is to seek a true completion of gravity, which would remain weakly coupled above $\Lambda_\text{UV}$. The consistency of the gravitational S-matrix requires higher-spin towers of species \cite{Camanho:2014apa}, whose dynamics is constrained \cite{Geiser:2022exp, Arkani-Hamed:2023jwn, Cheung:2024obl} to have stringy signatures. The consistency of black-hole thermodynamics at $\Lambda_\text{UV}$ also supports this picture \cite{Basile:2023blg, Bedroya:2024ubj, Herraez:2024kux}. Beside the above motivations for a small gravitational cutoff, the smallness of dark energy points to the same conclusion via holographic entropy bounds \cite{Castellano:2021mmx, Calderon-Infante:2023ler} as well as the considerations in \cite{Lust:2019zwm}, supported by the arguments in \cite{Stout:2022phm}. All in all, we are led to investigate weakly coupled strings in four-dimensional weakly curved asymptotically flat spacetime, which (at least at weak coupling) is the only option compatible with cosmology \cite{Dvali:2024dlb}. Alternatives would involve a non-perturbative UV completion such as M/F-theory, and compactifications thereof. \\

\emph{Setup. ---} We focus on closed strings for simplicity \footnote{Open strings are an interesting extension for future work \cite{Cribiori:2020sct, Leone:2023qfd}.}, postponing comments on additional contributions to the next section. This setting allows us to leverage the powerful framework of worldsheet conformal field theory (CFT). The spacetime sector is described by (super-)conformal characters of the spacetime isometries, while the internal degrees of freedom are captured by some unitary compact CFT whose central charges are fixed by criticality. The vacuum energy density at tree level in the string coupling $g_s$ \emph{vanishes by consistency}, an omen of \cref{eq:punchline}. The one-loop contribution in $d$ spacetime dimensions is given by $\Lambda_\text{dark}^{(1)} = - \, M_s^d \, \mathcal{I}$, with
\begin{eqaed}\label{eq:vac_torus_amplitude}
    \mathcal{I} = \int_\mathcal{F}d\mu \, Z(\tau, \overline{\tau}) \, .
\end{eqaed}
This yields the torus amplitude, obtained integrating the torus partition function $Z$ of the worldsheet CFT over a fundamental domain $\mathcal{F}$ of the modular group $\text{PSL}(2,\mathbb{Z})$, parametrized by the Teichm\"{u}ller variable $\tau = x + iy$. The invariant measure is $d\mu = \frac{d^2\tau}{y^2}$.

The resulting finite quantity is a function of moduli of the internal CFT, if any, encoding a potential for scalar fields. On general grounds, the spacetime and internal sectors of the CFT can mix according to
\begin{eqaed}\label{eq:CFT_Z}
    Z(\tau, \overline{\tau} \,|\, t) = \sum_s Z_\text{ext}^{(s)} \, Z_\text{int}^{(s)}(t) \, ,
\end{eqaed}
where from now on we suppress the $\tau, \, \overline{\tau}$ arguments but keep the moduli $t$, if any, explicit. The external sectors $Z_\text{ext}^{(s)}$ bring along a prefactor $y^{1-\frac{d}{2}}$ arising from bosonic zero-modes of the (transverse) embedding fields $X^\mu$. To make the modular covariance of the internal sectors manifest, we keep all fermionic characters in $Z_\text{ext}^{(s)}$ and furthermore we divide $Z_\text{ext}^{(s)}$ and multiply $Z_\text{int}^{(s)}$ by $y^{\frac{c}{2}}$, where $c=10-d$ is the central charge captured by $Z_\text{int}^{(s)}$ \cite{Afkhami-Jeddi:2020hde, Aoufia:2024awo, Abel:2024twz}. Prototypical examples of the latter are lattice sums $Z_\Gamma = \sum_{(h,\overline{h})\in \Gamma} e^{2\pi i (h \tau - \overline{h} \overline{\tau})}$. These ``reduced'' partition functions enter the gravitational cutoff $\Lambda_\text{UV}$ according to the (regularized) one-loop expression
\begin{eqaed}\label{eq:UV_cutoff_1-loop}
    \Lambda_\text{UV}^{-6} = 2 \zeta(3) \, M_s^{-6} + 2\pi \, M_s^{d-8} \int_\mathcal{F} d\mu \, \sum_s b_s \, Z_\text{int}^{(s)}
\end{eqaed}
derived in \cite{Aoufia:2024awo} for RNS-RNS superstrings, generalizing well-known examples. The constants $b_s$ encode the local low-energy expansion of the four-graviton scattering amplitude.

In \cite{Aoufia:2024awo} it was shown that $\Lambda_\text{UV} \ll M_\text{Pl}$ requires the presence of a light tower of species. One way to achieve this is to send $g_s \to 0^+$, leading to a tower of light string excitations with $m_\text{gap} = \Lambda_\text{UV} = M_s$. In this case, $\Lambda_\text{dark} \sim M_s^d$ unless it vanishes. Without fine-tuning, this case is excluded by experiments. Alternatively \footnote{Even when a species limit is driven primarily by an internal modulus $t \gg 1$ we keep $g_s \lesssim 1$ for reliability \cite{Aoufia:2024awo, Ooguri:2024ofs}.}, light species arise from a small spectral gap $\Delta_\text{gap} = \frac{m_\text{gap}^2}{M_s^2} \ll 1$ of the internal CFT, which lies at infinite distance in the conformal manifold \cite{Ooguri:2024ofs} and can be achieved sending some modulus $\Delta_\text{gap}^{-1} \equiv t \gg 1$ to infinity. Moreover, under some technical assumptions on the CFT \cite{Aoufia:2024awo, Ooguri:2024ofs}, in \emph{any such limit} at least part of the degrees of freedom of the CFT appear as $p \leq c$ extra dimensions which decompactify, with $m_\text{gap} = m_\text{KK} = \frac{M_s}{\sqrt{t}}$ and $\Lambda_\text{UV} = m^{\frac{p}{d+p-2}}$. In \cite{Aoufia:2024awo}, this result was obtained studying the modular properties of \cref{eq:UV_cutoff_1-loop} for $t \gg 1$. For convenience, if $p < c$ we absorb the remaining internal CFT sectors into the $Z_\text{ext}^{(s)}$, which now comes with a prefactor $y^{1-\frac{d+p}{2}}$ to the sum over states. \\

\emph{From the UV to the IR. ---} The upshot of the analysis of \cite{Aoufia:2024awo} is twofold. Firstly, the reduced internal partition functions satisfy the asymptotic differential equation
\begin{eqaed}\label{eq:UV_modular_diff_eq}
    \left(-t^2 \partial_t^2 - (2-p) t \partial_t + w_p\right) Z_\text{int}^{(s)} \sim \Delta_\tau \, Z_\text{int}^{(s)} \, ,
\end{eqaed}
where $\Delta_\tau \equiv -y^2(\partial_x^2 + \partial_y^2)$ is the Laplace-Beltrami operator on $\mathcal{F}$ and $w_p \equiv \frac{p}{2}\left(1-\frac{p}{2}\right)$. Secondly, from \cref{eq:UV_modular_diff_eq} and the spectral theory of $\Delta_\tau$ it follows that
\begin{eqaed}\label{eq:int_spectral_decomp}
    Z_\text{int}^{(s)} \sim t^{\frac{p}{2}} \left( A_s + F_s(t) \right)
\end{eqaed}
as $t \gg 1$, with the leading term $A_s$ independent of $t$ and $F_s \ll 1$ pointwise in $\tau$. The leading behavior is the volume (in string units) of the emerging extra dimensions, and from \cref{eq:vac_torus_amplitude} the leading behavior is $\mathcal{I}(t) \sim A \, t^{\frac{p}{2}}$. From \cref{eq:UV_modular_diff_eq} and \cref{eq:int_spectral_decomp}, one finds
\begin{eqaed}\label{eq:corr_modular_diff_eq}
    \left(-t^2 \partial_t^2 - 2 t \partial_t \right) F_s \sim \Delta_\tau \, F_s \, ,
\end{eqaed}
so that the correction $\delta \mathcal{I}$ to $\mathcal{I}(t) \equiv t^{\frac{p}{2}} \left(A + \delta \mathcal{I} \right)$ satisfies
\begin{eqaed}\label{eq:corr_integrated_diff_eq}
    \left(-t^2 \partial_t^2 - 2 t \partial_t\right) \delta \mathcal{I} \sim \int_\mathcal{F} d\mu \, \sum_s F_s \, \Delta_\tau Z_\text{ext}^{(s)} \, .
\end{eqaed}
In the $t \gg 1$ limit the leading contribution to the right-hand side of \cref{eq:corr_integrated_diff_eq} comes from the external vacuum, which amounts to the prefactor $y^{1-\frac{d+p}{2}}$. Since $\Delta_\tau y^{1-\frac{d+p}{2}} = w_{d+p} \, y^{1-\frac{d+p}{2}}$, one obtains
\begin{eqaed}\label{eq:IR_diff_eq}
    \left(-t^2 \partial_t^2 - 2 t \partial_t\right) \delta \mathcal{I} \sim w_{d+p} \, \delta \mathcal{I} \, ,
\end{eqaed}
whose general solution is a linear combination of $t^{-\frac{d+p}{2}}$ and $t^{\frac{d+p}{2}-1}$. The latter cannot contribute since $F_s \ll 1$. All in all, we obtain the one-loop expression
\begin{eqaed}\label{eq:final_1-loop_vac}
    \Lambda_\text{dark}^{(1)} & = M_s^d \, t^{\frac{p}{2}} \left(a_1 + b_1 \, t^{-\frac{d+p}{2}} \right) \\
    & = a_1 \, M_s^d \left( \frac{M_s}{m_\text{KK}}\right)^p + b_1 \, m_\text{KK}^d
\end{eqaed}
as $t \gg 1$, up to exponentially suppressed corrections due to massive string states. In particular, the constant $a_1$ is the \emph{higher-dimensional} torus amplitude \cite{Abel:2024twz}. Thus, the case $a_1=0$ has special significance, for instance tied to supersymmetry restoration \cite{Montero:2022prj, Anchordoqui:2023oqm} in $d+p$ dimensions. This occurs \emph{e.g.} in Scherk-Schwarz compactifications. The case $a_1 \neq 0$, which occurs \emph{e.g.} in the $\text{O}(16) \times \text{O}(16)$ heterotic model \cite{Alvarez-Gaume:1986ghj, Dixon:1986iz}, remains non-supersymmetric upon decompactification, bringing along several dynamical subtleties \cite{Basile:2018irz, Antonelli:2019nar, Raucci:2023xgx}.

\section{Consequences for phenomenology}\label{sec:pheno}

\subsection{Dark dimension: one-loop vacuum energy}\label{sec:sub_DD}

Let us examine the physical consequences of \cref{eq:final_1-loop_vac} more closely. We now specialize to $d=4$ and $\Lambda_\text{dark} \approx 10^{-120}$. To begin with, the smallness of dark energy requires that $a_1=0$,
\emph{i.e.} the absence of the one-loop higher-dimensional vacuum energy. Since $m_\text{KK} \lesssim M_s$ by spectral-gap bounds \cite{Hellerman:2009bu, Hellerman:2010qd}, the alternative $a_1\neq0$ with $M_s = \Lambda_\text{dark}^{\frac{1}{4}} \approx 10 \text{ meV}$ is incompatible with the observational constraint $M_s \gtrsim 10 \text{ TeV}$ \footnote{A fine-tuning with $t = \mathcal{O}(1)$ generally incurs string-scale forces, and is radiatively unstable.}. Therefore, \cref{eq:final_1-loop_vac} reduces to \cref{eq:punchline} with $m_\text{gap} = m_\text{KK}$ \footnote{In ``super no-scale'' settings \cite{Kounnas:2016gmz, Abel:2017rch} $a_1=b_1=0$, but the corrections $\Lambda_\text{dark} \lesssim M_s^4 \, e^{- \frac{M_s^2}{m_\text{KK}^2}}$ are overwhelmed by higher loops.}, the scale of $p$ emerging dimensions \cite{Aoufia:2024awo, Ooguri:2024ofs}.

From the above considerations, writing $b_1 \equiv \lambda^4$ we obtain
\begin{eqaed}\label{eq:final_1-loop_vacDD}
    \Lambda_\text{dark} \sim \lambda^4 \, m_\text{KK}^4 \, .
\end{eqaed}
This asymptotic relation was obtained by swampland considerations and combined with observational data in \cite{Montero:2022prj}, where it was shown that $p=1$ and the resulting mesoscopic extra dimension is roughly of micron size. In particular, the lower bound of $m_\text{gap}^d$ for dark energy was argued for by a one-loop Casimir energy estimate, which would have to be finely tuned to avoid the window $2 \leq \alpha \leq d$ \cite{Rudelius:2021oaz} in \cref{eq:scaling_relation}. 
Here we recover precisely $\alpha = d$, which was shown in \cite{Montero:2022prj} to be the unique option compatible with observations. In contrast, compactifications without scale separation have $\alpha = 2$. In \cite{Montero:2022prj} the prefactor $\lambda$ in \cref{eq:final_1-loop_vacDD} was restricted to the range $\lambda \approx 10^{-1} \divisionsymbol 10^{-3}$ by experimental bounds. In this case $M_s = \Lambda_\text{UV}$ is the five-dimensional Planck scale 
\begin{eqaed}\label{eq:DD_cutoff}
    \Lambda_\text{UV} = m_\text{KK}^{\frac{1}{3}} \sim \lambda^{-\frac{1}{3}} \, \Lambda_\text{dark}^{\frac{1}{12}}\approx 10^9 \divisionsymbol 10^{10} \text{ GeV} \, .
\end{eqaed}
For this relation, the string coupling constant $g_s$ is assumed to be of order one. Smaller values for $g_s$ would lead to a lower string scale $M_s$. In fact,
the dark dimension scenario is compatible with SM gauge couplings $g_s \approx g_\text{YM}^2$ on D-branes. 

Furthermore, this scenario is particularly appealing for the simple potential explanations of dark matter \cite{Anchordoqui:2022txe, Gonzalo:2022jac, Anchordoqui:2022tgp}, the electroweak hierarchy \cite{Montero:2022prj} and supersymmetry breaking \cite{Anchordoqui:2023oqm} (if any). The emergence of this scenario from worldsheet modular invariance places it on a firmer theoretical footing from the vantage point of string theory. \\

\emph{Additional contributions from branes. 
---} For SM gauge couplings to be acceptable, the SM degrees of freedom ought to be localized on D-branes \cite{Montero:2022prj, Heckman:2024trz}. Thus, one might worry about their tension contributing to dark energy. Indeed, D$q$-branes and O$q$-planes filling spacetime and wrapping a $(q-3)$-dimensional internal cycle $\Sigma$ contribute
\begin{eqaed}\label{eq:D-brane_O-plane_contribution}
    \abs{\Lambda_{\text{D}q/\text{O}q}} = g_s^{-1} \, M_s^{q+1} \, \text{Vol}(\Sigma) \gtrsim M_s^4
\end{eqaed}
regardless of internal anisotropies \cite{Hellerman:2009bu, Hellerman:2010qd} (see also \cite{FierroCota:2023bsp}). The same applies to NS5-branes, whose tension is given by $g_s^{-2} \, M_s^6$. Since experimentally $M_s \gtrsim 10 \text{ TeV}$, these are unacceptably large, even with a three-loop suppression. Hence, one must concoct tensionless combinations of D-branes and (negative-tension) O-planes or strong internal warping \cite{Blumenhagen:2022zzw}. Such configurations can occupy different regions in the internal dimension(s) to realize grand unification \cite{Heckman:2024trz}, favoring orthogonal or symplectic groups such as $\text{SO}(10)$ \footnote{Unitary groups with O-planes may also be possible \cite{Sagnotti:1995ga}.}.

\subsection{Finely tuned scenarios: higher loops and little strings}\label{sec:LST}

We now examine higher-loop corrections and alternative scenarios, which require some additional fine-tuning.
The genus expansion of the $(d+p)$-dimensional closed-string vacuum energy, reduced to $d$ dimensions, leads to the effective action
\begin{eqaed}\label{eq:eff_action_d+ploop}
    S_\text{eff} & \supseteq  \int d^dx \, \sqrt{-g} \left( M_\text{Pl}^{d-2} \, R - \sum_{g=1}^\infty \Lambda_{d+p}^{(g)} \left( \frac{M_s}{m_\text{KK}}\right)^p \right)
\end{eqaed}
with $M_\text{Pl}^{d-2} = g_s^{-2} \, M_s^{d+p-2} \, m_\text{KK}^{-p}$ and $\Lambda_{d+p}^{(g)} = a_g \, g_s^{2g-2} \, M_s^{d+p}$. In $d=4$, this contribution to the $g$-loop vacuum energy is then
$a_g \, g_s^{2g-2} \, M_s^{4+p} \, m_\text{KK}^{-p}$. Similarly, we estimate that the $g$-loop Casimir contribution also scale as $m_\text{KK}^4$ \cite{vonGersdorff:2005ce, Abel:2017rch}, such that \cref{eq:final_1-loop_vac} generalizes to
\begin{eqaed}\label{eq:final_g-loop_vac}
    \Lambda_\text{dark}^{(g)} 
    = g_s^{2g-2} \left(a_g \, M_s^4 \left(\frac{M_s}{m_\text{KK}}\right)^p + b_g \, m_\text{KK}^4\right) .
\end{eqaed}
It is easy to see that for the dark dimension scenario, with $g_s \approx 10^{-1} \divisionsymbol 10^{-3}$, all loop contributions to the first term must vanish up to a very high genus, \emph{i.e.} $a_g=0$ for $g < 20 \divisionsymbol 50$ or so. This essentially means that higher-dimensional supersymmetry must be unbroken, barring an enormous fine-tuning.

Therefore, we now investigate alternatives with  $m_\text{KK} \lesssim M_s$ and smaller string coupling $g_s$. Since $a_1 \neq 0$ is incompatible with observations, 
the leading vacuum energy must appear at higher orders \cite{Kachru:1998hd, Kachru:1998pg, Angelantonj:2004cm, Abel:2017rch}.

\emph{Two-loop contribution. 
---} Similarly to the one-loop case, $a_2 \neq 0$ is also excluded by observations, since it would lead to too low a string scale $M_s$. Therefore, at two loops we can only allow $b_2\neq0$ and $\Lambda_\text{dark} \sim g_s^2 \, m_\text{KK}^4$ once more requires $p=1$. But now $m_\text{KK} = g_s^{-\frac{1}{2}} \, \Lambda_\text{dark}^{\frac{1}{4}}$ varies with $g_s$. Since $M_s = g_s^{\frac{1}{2}} \, \Lambda_\text{dark}^{\frac{1}{12}}$ and $g_s^2 \, M_\text{Pl}^2 = M_s^{2+p} \, m_\text{KK}^{-p} \gtrsim M_s^2$, the allowed range for $g_s$ turns out to be $\Lambda_\text{dark}^{\frac{1}{12}} \approx 10^{-10} \lesssim g_s \lesssim 1$. This translates into $\Lambda_\text{dark}^{\frac{1}{4}} \lesssim m_\text{KK}\lesssim \Lambda_\text{dark}^{\frac{5}{24}}$, where the upper bound corresponds to a dark dimension  of size $m_\text{KK}^{-1}$ five orders of magnitude smaller than a micron. $M_s$ can now vary between $\Lambda_\text{dark}^{\frac{1}{8}} \lesssim  M_s\lesssim\Lambda_\text{dark}^{\frac{1}{12}}$. The lower bound is $E_\text{acc} \approx 10 \text{ TeV}$, the relevant scale for collider experiments. However, small string couplings $g_s \ll 1$ are incompatible with the SM, and NS5-branes with $g_s$-independent couplings must be introduced \cite{Antoniadis:2001sw}. This entails the additional fine-tuning problem of suppressing their contribution to dark energy.

\emph{Three-loop contribution. 
---} If the leading vacuum energy arises at three loops, a non-vanishing $a_3$ is now allowed by the experimental limits: namely for $a_3 \neq 0$ and any $p$ the relation $\Lambda_\text{dark} \sim g_s^4 \, M_s^4$ fixes 
\begin{eqaed}\label{eq:LST_scaling}
\Lambda_\text{UV} = M_s = m_\text{KK} = \Lambda_\text{dark}^{\frac{1}{8}} = E_\text{acc} \approx 10 \text{ TeV} \, .
\end{eqaed}
In other words, all scales are set by a string scale at the edge of detectability. This setting is the ``little string theory'' scenario of \cite{Antoniadis:2001sw}, here rediscovered from the cosmological hierarchy. Once more, $g_s = \Lambda_\text{dark}^{\frac{1}{8}}
\approx 10^{-15}$ cannot provide SM gauge couplings without NS5-branes. 

On the other hand, for $a_3 = 0$ and $b_3\neq0$, 
 \cref{eq:LST_scaling} follows also  if $p > 2$. The case $p=2$ fixes $M_s = E_\text{acc}$ but not $g_s$, whereas $p = 1$ \emph{interpolates} between \cref{eq:DD_cutoff} and \cref{eq:LST_scaling} upon varying $g_s$ in the allowed range
$ \Lambda_\text{dark}^{\frac{1}{8}}
\approx 10^{-15} \lesssim g_s \lesssim 1$.

Note that, assuming that the Casimir energy arise only at three loops, one can naturally identify $\lambda$ in \cref{eq:final_1-loop_vacDD} with $g_s$, and the required value $\lambda \approx 10^{-1} \divisionsymbol 10^{-3}$, obtained from gravitational bounds, also leads to the correct values for SM gauge couplings on D-branes.

\begin{figure}[ht!]
    \centering
    \includegraphics[scale=0.35]{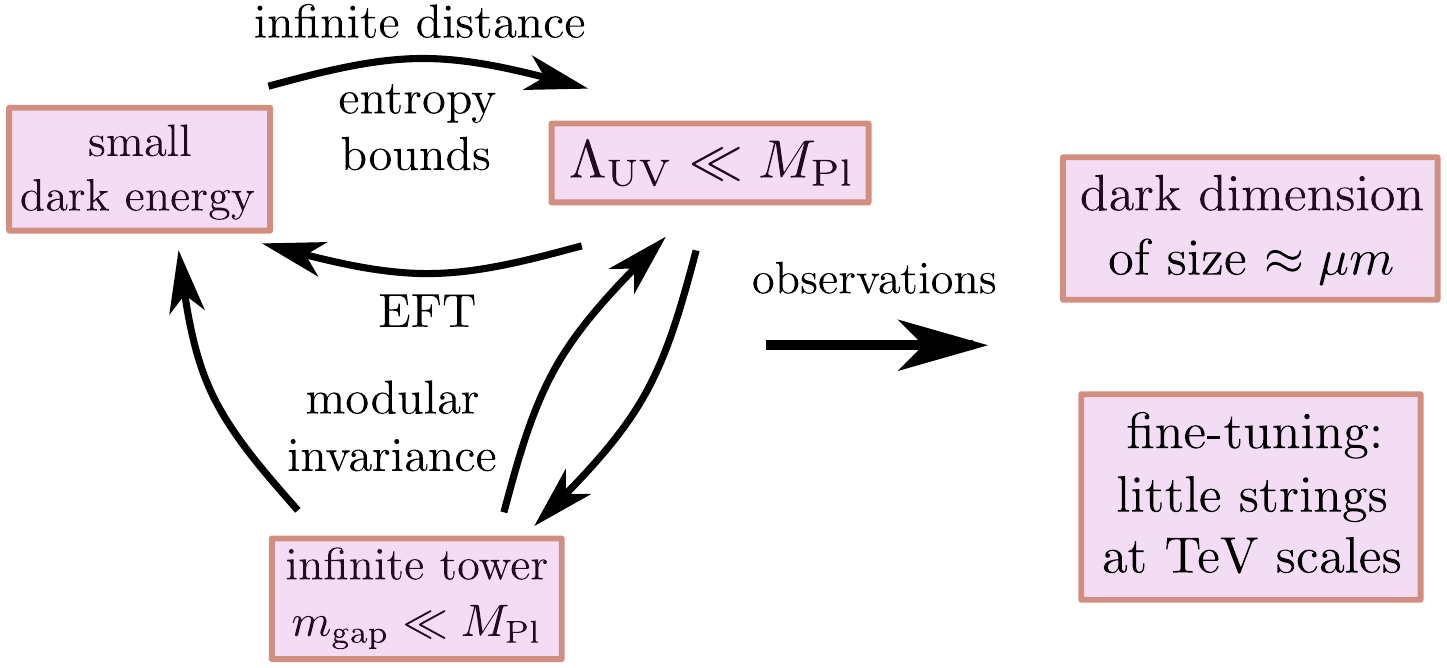}
    \caption{A diagram of connections between relevant scales and the phenomenological scenarios they led to.}
    \label{fig:diagram}
\end{figure}

\section{Conclusions}\label{sec:conclusions}

In this letter we have discussed the emergence of the relation in \cref{eq:scaling_relation} from the consistency of string theory with observational data. The smallness of dark energy is expected to be accompanied by towers of light species \cite{Lust:2019zwm, Stout:2022phm}. We showed that these arise for $\Lambda_\text{UV} \ll M_\text{Pl}$, and are either the Kaluza-Klein modes of the dark dimension scenario of \cite{Montero:2022prj} or, with some fine-tuning, higher-spin excitations in the little string theory scenario of \cite{Antoniadis:2001sw}. This result resonates well with the emergent string conjecture \cite{Lee:2019wij}, which states that only two kinds of infinite-distance limits in quantum gravity are decompactifications or tensionless string limits (up to dualities). In the context of weakly coupled closed strings, this dichotomy was shown in \cite{Aoufia:2024awo} and \cref{eq:punchline} arises from the modular properties of the internal CFT, at least given the technical assumptions of \cite{Aoufia:2024awo, Ooguri:2024ofs} on small-gap limits. The tower of species which accompanies a small UV cutoff is assumed to become uniformly light, although the conclusions are expected to hold more generally \cite{Ooguri:2006in, Stout:2022phm}. As a result, observational data selects the dark dimension scenario of \cite{Montero:2022prj} as the \emph{most natural viable option}. 

The connection between IR and UV scales relates new physics with the mysteries of dark energy in a hallmark of UV/IR mixing, as depicted in \cref{fig:diagram}, and resonates with holography via the entropy bound $\Lambda_\text{UV} \lesssim \Lambda_\text{IR}^{\frac{1}{d-1}}$ applied to dark energy $\Lambda_\text{dark} \equiv \Lambda_\text{IR}^d$. In fact, the relation $\Lambda_\text{dark} \sim m_\text{KK}^d$ satisfies it for any number of extra dimensions, and the scalings in the dark dimension scenario \emph{saturate it}. The interconnected web of consistency constraints uncovered by the swampland program thus manifests in our setting via modular invariance, and the smallness of dark energy turns into a predictive advantage. Indeed, due to the UV/IR mixing embodied by \cref{eq:punchline} --- which is invisible in EFT \cite{Branchina:2023ogv, Anchordoqui:2023laz, Branchina:2024ljd} --- it leads to experimentally accessible and surprisingly constrained features of new physics. 

Although the results that we presented in this letter rely on certain genericity assumptions, and could in principle be circumvented in some finely tuned settings which we discussed, the overall message remains: this connection between deep theoretical principles and phenomenological possibilities opens an exciting window into near-future experimental tests of string theory and possibly quantum gravity in a broader sense.

\section*{Acknowledgements}
We thank L. Anchordoqui, I. Antoniadis, A. Herr\'{a}ez, M. Montero, L. Nutricati, A. Platania, S. Raucci and C. Vafa for feedback and discussions, and C. Aoufia and G. Leone for collaboration on related projects. The work of I.B. is supported by the Origins Excellence Cluster, and the work of D.L. is supported by the Origins Excellence Cluster and the German-Israel-Project (DIP) on Holography and the Swampland.

\bibliographystyle{apsrev4-1}
\bibliography{refs}

\end{document}